\documentclass[10pt]{article}
\usepackage{a4wide}
\title{{\bf Quantum-wave pattern recognition: \\
from simulations towards implementation}}
\author{Mitja Peru\v s \thanks{Corresponding author:
perus@icg.tu-graz.ac.at}
\ and Horst Bischof \\
Graz University of Technology, Institute for Computer Vision and
Graphics
\\ Inffeldgasse 16, 2.OG, A-8010 Graz, Austria \\
www.icg.tu-graz.ac.at/$\sim$perus \& $\sim$bischof}
\date{}
\twocolumn 
\begin{document}
\maketitle
\begin{abstract}
A neural-net-like model, which is realizable using quantum
holography, is proposed for quantum associative me\-mory and
pattern recognition. This Hopfield-based mathematical
model/algorithm, translated to quantum formalism, has been
successfully tested in computer simulations of concrete
pattern-recognition applications. In parallel, the same
mathematics governs quantum dynamics which can be harnessed for
information processing by proper (de)coding manipulation. Since we
are able to give quantum interpretation to all the elements (e.g.,
variables, couplings) of the model, and as far as we are able to
show that processing, governed by that mathematics, is
experimentally implementable in real quantum systems, we can
expect efficient quantum computing -- in our case pattern
recognition based on quantum content-addressable
associative memory. 
\normalsize \\
\\
\end{abstract}
\thispagestyle{empty}
{\it 1. INTRODUCTION} \\

Quantum neural nets \cite{kasabov,nq} are a branch of quantum
computers needing no logic gates. It will be shown that the
implementation of associative neural nets can be {\it
naturally}-physical, i.e. no artificial classical-physical devices
are necessary, except for encoding into and decoding from quantum
systems. To demonstrate quantum implementation of certain ANN, it
is best to remember a fundamental technique which has already been
much experimentally tested and used -- holography. Holography is a
practical 3D image-storage and -reconstruction procedure
\cite{hol}. Its imaging is powerful and of high resolution,
although the technique is relatively simple --- it uses merely
reflection from the laser-illuminated object and interference of
the "object"-beam with a "reference"-beam. Early associative
memories were inspired by holography \cite{early}. They were a
version of digitalized amplitude-information holography. Merging
of ANN and holograph\-ic approach \cite{hnet,ron} continues to be
very useful.

Since holography can be in principle realized applying any sort of
coherent waves \cite{hol}, it is realizable also using quantum
waves or wave-packets. The latter are of the same type as the
Gabor wavelets used, e.g., in computer vision \cite{wavelet}.
Dennis Gabor got the Nobel prize for physics for his invention of
holography. The fast-developing nanotechnology promises to realize
it soon in quantum field, i.e. using quantum
probability-distribution waves \cite{qhol}.

This paper introduces some relations of parallel-distri\-buted
processes (PDP) in neural and quantum systems (overview in
\cite{ana}), and their relations to holography, in order to
propose a {\it quantum pattern recognition} model. The mathematics
of this algorithmic model to be presented has already been
successfully computer-simulated, e.g. in \cite{simul}. The next
step made here is to proceed from simulations to a proposal of
quantum implementation, supported by considering quantum-optical /
-holographic experiments.

This step was enabled by the following observation on the {\it
parallel-distributed information-encoding into waves}, i.e. into $
(A_1 e^{i \varphi_1}, A_2 e^{i \varphi_2}, ..., A_N e^{i
\varphi_N}) $, in general. It has two special cases: (I) encoding
in amplitudes $A$ only, i.e. in $ (A_1, A_2, ..., A_N) $, and (II)
encoding in oscillatory phases $\varphi$ only, i.e. in $ (e^{i
\varphi_1}, e^{i \varphi_2},..., e^{i \varphi_N}) $, $i =
\sqrt{-1}$. These two cases enable {\it effectively the same
information processing} as far as the following variable exchange
can be made in the mathematics of the model/algorithm: $ A
\leftrightarrow e^{i \varphi} $. It will be shown for our
wave-model (II) with $A=1$: {\it what works for real-valued
coding-numbers (A) works for sinusoid encoding also}. I.e.,
"wave-based" model (II) is equivalent to "intensity-based"
model (I).

In sec. 2 we "translate" Hopfield's model into quantum formalism
\cite{nq,ana}. In sec. 3 we "transform" Hopfield's model into
wave-model (II), showing their {\it equivalence} for pattern
recognition. In sec. 4 we present possible quantum implementations
of the simulated model. \\
\\
{\it 2. QUANTUM-WAVE HOPFIELD-BASED NET} \\

The simplest Hopfield ANN (1982) incorporates Hebbian
memory-storage "into" correlation matrix {\bf J}, i.e. ${\bf J} =
\sum_{k=1}^P \vec{v}^k \otimes \vec{v}^k$ ($\otimes$ denotes
tensor/outer product), and a memory-influenced transformation of
patterns $\vec{v}$: $ \vec{v}^{output} = {\bf J} \vec{v}^{input}
$. Each of $P$ patterns, simultaneously stored in the same
net/{\bf J}, is denoted by a superscript index $k$: $k=1, ..., P$.
Patterns $\vec{v}^k$, which become Hopfield-net's eigenstates
(attractors), can be complex-valued and can be quantum-encoded (as
will be shown). Therefore, we will henceforth use the quantum
notation, $\psi^k$, for them (i.e., $\vec{v}^k = \psi^k$, instead
of the standard case: $\vec{v}^k = A^k$). (So-called wave-function
$\Psi$ describes the whole state of the quantum system/net;
$\psi^k$ describes the $k^{th}$ of its eigenstates.) Thus,
patterns are assumed to be encoded into quantum
eigen-wave-functions $\psi^k$ (physical realization will be
discussed later).

Turning from global description of the quantum PDP (using
associative memory {\bf J} and net-states $\Psi$) into local one
(using interaction-weights $J_{hj}$ and unit-states $\Psi_j$;
$j,h=1,...,N$, for $N$ units; $N$ is huge), we have:
$$ J_{hj} = \sum_{k=1}^P \psi_h^k (\psi_j^k)^{\ast}
\eqno(1) ,
$$ where the asterisk denotes complex conjugation, and
$$ \Psi_h^{output} = \sum_{j=1}^N J_{hj} \Psi_j^{input}
\eqno(2) . $$
Inserting eq. (1) into eq. (2), we obtain:
$$ \Psi_h^{output} = \sum_{j=1}^N
\left( \sum_{k=1}^P \psi^k_h (\psi^k_j)^{\ast} \right)
\Psi_j^{input} = $$
$$ = \sum_{k=1}^P \left( \sum_{j=1}^N (\psi_j^k)^{\ast} \Psi_j^{input}
\right) \psi_h^k = \sum_{k=1}^P c^k \psi_h^k \eqno(3) . $$
Usually, exclusively those coefficient $c^k$, say $c^{k_0}$, is
close to 1 which belongs to the memorized pattern $\psi^{k_0}$
which is the most similar to $\Psi^{input}$. Consequently, all
other $c^k$, $k \neq k_0$, are close to zero. In such a case of
the process of eq. (3), the quantum associative net {\it
recognizes} the input-pattern. See analysis of this matching
process in \cite{nq,q}.

{\bf J} is called Green-function propagator, $G$, in quantum
theory \cite{ana,q}. $G$'s description of input--output
transformations corresponds to statistical description of
state-relations (relations in encoded data) by the quantum {\it
density matrix} {\bf $\rho$}. We will not use $\rho$ here; we
merely wanted to emphasize $\rho$'s role as a "quantum archive"
(of all potential input--output transformations, in contrary to
$G$'s actual ones).

For physics-oriented readers, we write eq. (2), with kernel of eq.
(1) inserted, into space-time form: $$ \Psi (\vec{r}_2, t_2) = $$
$$ = \int \int \left( \sum_{k=1}^P \psi^k (\vec{r}_2, t_2)
(\psi^k (\vec{r}_1, t_1))^{\ast} \right) \Psi (\vec{r}_1, t_1)
d\vec{r}_1 dt_1 \eqno(4) . $$ We replaced unit-indices $h, j$ by
($\vec{r_2}, t_2$), ($\vec{r_1}, t_1$), and discrete summation by
an integration over the whole effectively-continuous quantum
system/net (if it consists of very many "units"). This is the
 Feynman (path-integ\-ral) version of the Schr\"{o}dinger
equation, the fundamental equation for quantum dynamics --- in
Dirac's notation: \\ $ \mid \Psi \rangle = \mid \Psi \rangle
\langle \Psi \mid \Psi \rangle = ( \sum_k \mid \psi^k \rangle
\langle \psi^k \mid ) \mid \Psi \rangle $.

The Hopfield computational model, incorporating coupled eqs. (1)
\& (2) with real-valued variables, has been used in very many
different applications of numerous authors. Based on \cite{early},
it is a historical prototype-model, out from which so many other
models, more applicable for particular problems, have been
developed. Using it, the first author has computationally
recognized patterns of approximated 3D structures of proteins
using a huge memorized data-base (from the Brookhaven protein data
bank) \cite{simul}. However, for quantum implementation of
associative PDP, we should first turn to this model, eqs. (1) \&
(2), again using it as a "Rosetta stone". This might then enable
subsequent fantastic improvements which are promised by
possibly-entangled \cite{entangl} quantum field dynamics
manipulated by so-called classical--quantum interactions. So, the
quantum breakthrough for ANN-implementations can best be made with
the prototypical associative content-addressable memory of eqs.
(1) \& (2), because its dynamics is {\it relatively similar to
natural processes}, mainly in spin systems (i.e., spin glass)
\cite{spin} and quantum
fields \cite{q}. \\
\\
{\it 3. SINUSOID ACTIVITIES OF NET'S "UNITS"} \\

Quantum wave-functions $\psi^k$ can have many forms. For our
purposes, (quantum-optical) plain-waves $ \psi^k (\vec{r}, t) =
A^k (\vec{r}, t) e^{i \varphi^k (\vec{r}, t)} $ are the most
appropriate. An advanced alternative, to be left for our future
work, are quantum wave-packets nearly-identical to Gabor wavelets
\cite{wavelet}.

Holography shows, at least for non-quantum waves, how one can
parallel-distributively encode patterns $k$ into a web of waves $
(A^k_1 e^{i \varphi^k_1}, A^k_2 e^{i \varphi^k_2}, ..., A^k_N e^{i
\varphi^k_N}) $. The amplitude $A_j^k$ and the oscillatory phase
$\varphi_j^k$ have the same lower index $j$ ($j=1,...,N$; $N$
huge), since they belong to the same "waving" point, which is our
"unit" (encoding a point of the pattern).

We {\it can} use plain-waves (sinusoids) with the {\it same}
constant amplitude, say $A=1$; so, $A^k_j = 1$ for all $k, j$.
This is functioning for few decades, known as phase-information
holography. We thus replace all $\psi$-variables $(\equiv v)$ in
eqs. (1), (2), (3) and (4), with $e^{i \varphi}$, instead of
Hopfield's $A$. We are allowed to do this -- it's a usual
mathematical exchange of variables. The essential observation is
that with this legal variable-exchange, $ A \leftrightarrow e^{i
\varphi} $, giving $\psi^k_j = e^{i \varphi^k_j}$ instead of
$\psi^k_j = A^k_j$, all the simulation-tested mathematics remains
valid for sinusoid-encoded patterns also. Thus, we can claim that
the Hopfield algorithm, i.e. eqs. (1) \& (2), {\it works with
complex-valued sinusoid-inputs at least as much as with
real-valued inputs}! Performance of the wave-phase model (II) with
eigenpatterns $\psi^k = e^{i \varphi^k}$, $A^k=1$, is {\it equal}
to performance of the amplitude model (I) with $\psi^k = A^k$,
$A^k$ real number. However, when using {\it both} -- different
amplitudes {\it and} different phases -- performance might be
(much) improved, as practically proved by HNeT \cite{hnet}. Much
better results arise using HNeT's preprocessing method \cite{hnet}
where inputs $v^k_j$ are {\it sigmoidally mapped} into phases
$\varphi^k_j$ to obtain a convenient symmetric (uniform)
data-distribution: $\varphi^k_j = 2\pi \left( 1+exp (
\frac{\bar{v}^k - v^k_j}{\sigma (v^k)} ) \right)^{-1}$.

To prove quantum-wave pattern recognition with the system of eqs.
(1) \& (2), it suffices to execute the exchange, $ \psi_j^k
\leftrightarrow e^{i \varphi_j^k} $, first in the Hebbian eq. (1),
using $ (e^{i \varphi})^{\ast} = e^{-i \varphi} $:
$$ G_{hj} = \sum_{k=1}^P e^{i \varphi_h^k} e^{-i \varphi_j^k} =
\sum_{k=1}^P e^{i (\varphi_h^k - \varphi_j^k)} \eqno(5) , $$ and
secondly in eq. (2). So, instead of eq. (2), when inserting now
expression (5) into $J_{hj}$ of eq. (2) and exchanging $ \psi_h^k
\leftrightarrow e^{i
\varphi_h^k} $ in it, we obtain the following
equivalent of eq. (3):
$$ e^{i \varphi_h^{output}} =
\sum_{j=1}^N \left( \sum_{k=1}^P e^{i \varphi_h^k} e^{-i
\varphi^k_j} \right) e^{i \varphi_j^{input}} = $$ $$ =
\sum_{k=1}^P \left( \sum_{j=1}^N e^{i \varphi_j^{input}} e^{-i
\varphi_j^k} \right) e^{i \varphi^k_h} \doteq e^{i
\varphi_h^{k_0}} \eqno(6).
$$

This enormous process of phase (mis)matching, producing
constructive or destructive interferences, is described in detail,
mathematically and informatically, in \cite{hnet} and for quantum
case in \cite{nq,q}. The right-most expression of eq. (6) really
describes the output, i.e. approximately $ e^{i \varphi_h^{k_0}}
$, {\it only if} the same conditions are valid as described below
eq. (3): If the input wave has a similar phase to one of the
memorized waves, say $k = k_0$, then those wave will be
reconstructed --- the pattern it is carrying, $k_0$, will be {\it
recognized}. See \cite{q,nq} for discussion of the precise
conditions for clear pattern-recall. If these conditions are not
satisfied by the data correlation-structure, interferences
("cross-talk") lead to a mixed or averaged output (details in
\cite{simul}).

So, instead of a long series of products (correlations) of {\it
real}-valued information-coding numbers, $A$, as in the Hopfield
model, we have here a long series of complex-valued {\it
exponentials} (waves) with differences of informati\-on-coding
phases, $\varphi$, in each exponent. These {\it phase-differences
(peak delays) encode discrepancies in data}. Our wave output $
e^{i \varphi^{k_0}_h} $ is the same as Hopfield's $A^{k_0}_h$. In
sum, {\it input--output transformations are the same} in the wave
case as it were in our simulated real-number (intensity) case
\cite{simul}. All this proves the pattern recognition capabilities
of the wave model (II) with phase-encoding of pattern-points $h$.
The memory is "represented" by the
hologram, i.e. wave-interference pattern, of eq. (5). \\
\\
{\it 4. DISCUSSION ON IMPLEMENTATION} \\

Mathematically and computationally \cite{simul}, we have {\it
proved the associative memory-storage and pattern-recog\-nition
performance} of Peru\v s's model named {\it Quantum Associative
Network} \cite{q}. It remains to prove it in quant\-um-physical
experimental practice, i.e. with real quantum pattern-encoding
waves, not merely with digital simulated ones (complex sinusoids).

It is crucial that our model \cite{q} is fundamental, optimized
and relatively {\it natural}, i.e. almost no artificial devices
are really necessary, in contrary to all other models. Laser is
also not unavoidable.

Implementation of our model is most appropriate using quantum-wave
holography which is within the reach of present experimental
technology \cite{qhol,entangl}. Several applied techniques, which
are at least partially quantum-holography-based, {\it are} already
functioning, e.g. some sorts of tomography (fMRI and PET scanning)
\cite{schempp}. Holo\-graphy is a fundamental and universal
procedure in the sense that, in principle, any sort of coherent
waves can be applied for interference-based simultaneous recording
of many objects into (and for selective reconstruction from)
various hologram-media. Apart of classical optical and acoustical
holography, microwave-, X-ray-, atom- and electron-holography were
realized \cite{hol}. There is just a step further to quantum-wave
holography as described here.

This attempt is supported by the following reports: According to
\cite{nielsen}, universal quantum computation is realized using
only projective measurement, like ours of eq. (3) or eq. (6),
quantum memory, like ours of eq. (1) or eq. (5), and preparation
of the initial state (the laser-wave in our case).
Information-storage and -retrieval through quantum phase
\cite{qphase} and measurements of quantum relative phase
\cite{relphase} have been experimentally demonstrated. Quantum
encodings in spin systems and coupled harmonic oscillators
are possible \cite{spin}, thus enabling Hopfield-like
pattern-storage and -recognition in such nets, including spin-wave
holographic ones.

However, if nanotechnology could not (which is unlikely) realize
quantum-holographic pattern-recognition as proposed here,
something like that is hypothesized to be happening in the
(visual) brain \cite{brain}. Not only brain, the whole quantum
Nature itself probably incorporates such processes, at least in
interaction with our quantum-measurement devices \cite{cahill}. In
worst case, it does merely not let us to collaborate with -- until
tomorrow?

Recently, we found similar quantum pattern-recogniti\-on proposals
\cite{qpr}. Trugenberger's one is related to the fact that a
special case of Hebbian memory-storage, eq. (1), i.e. with bipolar
states (1 and -1 only), is equivalent to quantum-implementable NOT
XOR gate. This makes a link between ANN-like and logic-gate-based
branches of quantum pattern recognition.

The benefits of the first branch, i.e. quantum neural-net
approach, are the following: {\it Quantum decoherence} ("collapse
of the wave-function") is {\it not devastating} (as is in
main-stream quantum computers), but is {\it usefully harnessed for
pattern recognition}. No special mechanisms are needed for quantum
error-correction, since it is done spontaneously by the net's
self-organizing process (as in ANN). Initialization problems are
not as serious \cite{Kak} as in logic-gate quantum computers, at
least not when an object is holographed. In this case, reflection
from the surface determines the phases, and fluctuations do not
destroy the modulation (cf., experimental quantum-phase storage
and retrieval \cite{qphase}). Finally, as it is characteristic for
quantum computers, quantum associative net is exponentially
superior to its classical counterparts in memory capacity,
processing speed and in miniaturization \cite{kasabov}. This
brings improvements in computational capacity and efficiency.
Quantum ANN promise to outperform logic-gate quantum computing in
associative tasks like discussed here, and in flexibility (fuzzy
processing) \cite{q}, where also classical ANN outperform
sequential computing. Finally, our net presented \cite{q} is
relatively inexpensive, because it is relatively natural, and is
of huge (at least) theoretical importance. \\
\\
\small {\it ACKNOWLEDGEMENTS:} \\ This work was enabled by the
European-Union's Marie-Curie fellowship for M.P (contract no.
HPMF-CT-2002-01808). M.P. is also grateful to Professors Karl H.
Pribram and Alexandr A. Ezhov for encouragement for fundamental
studies. 

\normalsize
\begin{thebibliography}{99}
\bibitem{kasabov} A.A. Ezhov, D. Ventura: Ch. 11 in N. Kasabov (ed.):
{\it Future Directions for Intelligent Systems and Information
Sciences}; Physica/Spinger, Heidelberg, 2000. [Cf.: E. Behrman et
al.: arXiv:quant-ph/0202131 \& Info. Sci. {\bf 128} (2000)
257-269.]

\bibitem{nq} M. Peru\v s: Neural Netw. World {\bf 10} (2000)
1001-13.

\bibitem{hol} E.g.: H. Bjelkhagen \& H. Caulfield (eds.):
{\it Selected Papers on the Fundamental Techniques in Holography};
SPIE Opt. Eng. Press, Bellingham (WA), 2001. \
 H. Caulfield (ed.): {\it Handbook of Optical
Holography}; Academic Press, New York, 1979. \ Or see (text)books
by: Hariharan; Stroke; Collier et al.; Saxby.

\bibitem{early} Willshaw, Buneman, Longuet-Higgins (1969); Kohonen (1972);
Hopfield (1982). Cf.: Nakano (1972); Anderson (1972); Kosko
(1988). [Early holographic associative memories: P. van Heerden,
1963; Gabor, 1969, (1949). Cf., Pribram, 1963, 1969.]

\bibitem{hnet} J. Sutherland: Int. J. Neural Sys. {\bf 1} (1990) 256-267.
\ Applications: www.ANDcorporation.com

\bibitem{ron} D. Psaltis et al.: Nature {343} (1990) 325-330. \ Sci.
Amer. {\bf 273}(5) (Nov. 1995) 52-58. \ R. Spencer in: C. Dagli et
al. (eds.): {\it Intelligent Engineering Systems Through ANN}
(vol. 10); ASME Press, St. Louis, 2000; pp. 971-6.

\bibitem{wavelet} T.S. Lee: IEEE Transac. Pattern Anal. \& Mach.
Intell. {\bf 18}(10) (1996) 1-13.  C. Chui: {\it An Introduction
to Wavelets}; Academic Press, Boston, 1992.

\bibitem{qhol} A. Abouraddy et al.: "Quantum holography";
Optics Express {\bf 9}(10) (2001) 498-505. \ B. Saleh et al. in:
{\it Proc. LEOS'96}, vol. 1, Laser \& El.Opt. Soc., Boston, 1996,
p. 362-3. \ Cf.: N. Bhattacharya et al.: Phys. Rev. Lett. {\bf 88}
(2002) 137901. [And: S. Takeuchi: Phys. Rev. A {\bf 62}(3) (2000)
032301. \ R. Spreeuw: Phys. Rev. A {\bf 63}(6) (2001) 062302.]

\bibitem{ana} M. Peru\v s: Nonlin. Phenom. in Complex Sys. {\bf 4} (2001)
157-193. \ Zeitschr. angew. Math. \& Mech. {\bf 78}, S 1 (1998)
23-26. \ Informatica {\bf 20} (1996) 173-183.

\bibitem{simul} M. Peru\v s: Int. J. Computing Anticip. Sys. {\bf 13} (2002)
376-391.

\bibitem{spin} S. Bartlett et al.: Phys. Rev. A {\bf 65} (2002)
052316. \ D. Lidar, O. Biham: Phys. Rev. E {\bf 56} (1997)
3661-81. \ [Cf.: NMR bulk / ensemble quantum computing: N.
Gershenfeld, I. Chuang: Science {\bf 275} (1997) 350-6. \ D. Cory
et al.: Proc. Nat. Acad. Sc. USA {\bf 94} (1997) 1634-9.]

\bibitem{q} M. Peru\v s, S.K. Dey: Appl. Math. Lett. {\bf 13}(8) (2000)
31-36. \ Orig.: M. Peru\v s in {\it Proc. JC-IS'98}, vol. 2:
$3^{rd}$IC-CI\&N; eds. P.P. Wang et al., N. Carolina, 1998, p.
197-200.

\bibitem{entangl} H. Lee at al.: Phys. Rev. A {\bf 65} (2002) 030101.

\bibitem{schempp} W. Schempp: "Quantum holography and MR tomography...";
Informatica {\bf 21} (1997) 541-562. \ Cf., G. D'Ariano et al.:
"Quantum tomography"; Phys. Lett. A {\bf 276} (2000) 25-30.

\bibitem{nielsen} M. Nielsen: arXiv:quant-ph/0108020.

\bibitem{qphase} J. Ahn et al.: Science {\bf 287} (2000) 463-5. \
Y. Wu: Phys. Rev. A {\bf 63}(5) (2001) 052303. \ Cf.: J. Denschlag
et al.: Science {\bf 287} (2000) 97-101.

\bibitem{relphase} A. Trifonov et al.: J. Optics B
{\bf 2} (2000) 105-112.

\bibitem{brain} M. Peru\v s: Informatica {\bf 25} (2001) 575-592. \
K.H. Pribram: {\it Brain and Perception}; LEA, Hillsdale (NJ),
1991. \ A.F. da Rocha et al.: Progress in Neurobiol. {\bf 64}
(2001) 555-573. \ R. Nobili: Phys. Rev. A {\bf 32} (1985) 3518-26.

\bibitem{cahill} Cf.: R. Cahill:
arXiv:quant-ph/0111026 or in {\it Proc. SPIE Conf. \#4590: BioMEMS
\& Smart Nanostructures}, ed. L. Kish, 2001, p. 319-328.

\bibitem{qpr} A.Yu. Vlasov: "Quantum computations and image recognition";
arXiv:quant-ph/9703010. \ C.A. Trugenberger: "Quantum pattern
recognition"; Phys. Rev. Lett. {\bf 89}(27) (2002) 277903. \ Phys.
Rev. Lett. {\bf 87}(6) (2001) 067901. \ arXiv:quant-ph/0210176. \
A. Ezhov, A. Nifanova, D. Ventura: Info. Sci. {\bf 128} (2000)
271-293. \ A. Ezhov in {\it Proc. ICAPR'01}, eds. Singh et al.,
Rio de Jan., p. 60-71.

\bibitem{Kak} S. Kak: Foundat. Phys. {\bf 29} (1999) 267-279.
\end{thebibliography}
\end{document}